\documentclass[10pt,  journal,  twocolumn]{IEEEtran}
\usepackage{amssymb}
\usepackage{graphicx}
\usepackage{amsmath}
\usepackage{epsf}
\usepackage{mathrsfs}
\usepackage{cite}
\newcounter{MYtempeqncnt}
\newtheorem{mytheo}{Theorem}
\newtheorem{corollary}{Corollary} 
\newtheorem{lemma}{Lemma}
\newtheorem{remark}{Remark} 
 
\begin{document}
\title{{On feasibility of perfect interference alignment in interference networks}}
\author{\Large Zainalabedin ~Samadi,  
       ~Vahid ~Tabatabavakili and ~Farzan ~Haddadi
\\\small Dept. of Elec. Eng.,   Iran University of Sceince and Technology 
Tehran,   Iran 
\\ \{z.samadi\}@elec.iust.ac.ir
\\ \{vakily,   haddadi\}@iust.ac.ir
}
\maketitle
\begin{abstract}
Interference alignment(IA) is mostly achieved by coding interference over multiple dimensions. Intuitively, the more interfering signals that need to be aligned, the larger the number of dimensions needed to align them. This dimensionality requirement poses a major challenge for IA in practical systems. This work evaluates the necessary and sufficient conditions on channel structure of a 3 user interference channel(IC) to make perfect IA feasible within limited number of channel extensions. It is shown that if only one of interfering channel coefficients can be designed to a specific value,   interference would be aligned perfectly at all receivers. 

\end{abstract}
\IEEEdisplaynotcompsoctitleabstractindextext
\section{Introduction}

One of the recent strategies to deal with interference is interference alignment. In a multiuser channel,   the interference alignment method puts aside a fraction of the available dimension at each receiver for the interference and then adjusts the signaling scheme such that all interfering signals are squeezed in the interference subspace. The remaining dimensions are dedicated for communicating the desired signal, keeping it free from interference. 

  Cadambe and Jafar \cite{Cadam08},   proposed the linear interference alignment (LIA) scheme for IC and proved that this method is capable of reaching optimal degrees of freedom of this network. The optimal degrees of freedom for a $K$ user IC is obtained in the same paper to be $K/2$. The proposed scheme in \cite{Cadam08} is applied over many parallel channels and achieves the optimal degrees of freedom as the signal-to-noise ratio (SNR) goes to infinity. Nazer et. al.  \cite{Nazer12},   proposed the so called ergodic interference alignment scheme to achieve 1/2 interference-free ergodic capacity of IC at any signal-to-noise ratio. This scheme is based on a particular pairing of the channel matrices. The scheme needs roughly the same order of channel extension compared with \cite{Cadam08},   to achieve optimum performance. The similar idea of opportunistically pairing two channel instances to cancel interference has been proposed  independently by \cite{Sang13} as well. However,   ergodic interference alignment scheme considers only an special pairing of the channel matrices and does not discuss the general structure of the paired channels  suitable for interference cancelation.

This paper addresses the general relationship between the paired channel matrices suitable for canceling interference,   assuming linear combining of paired channel output signals. Using this general pairing scheme,  to align interference at receiver, proposed scheme significantly lowers the required delay for interference to be canceled. 

From a different standpoint, this paper obtains the necessary and sufficient feasibility conditions on channel structure to achieve total DoF of the IC using limited number of channel extension. So far, Interference alignment feasibility literature have mainly focused on network configuration, see \cite{Ruan} and references therein. To ease some of interference alignment criteria by using channel structure,  \cite{Leejan09} investigates degrees of freedom for the partially connected ICs where some arbitrary interfering links are assumed disconnected. In this channel model,   \cite{Leejan09} examines how these disconnected links are considered on designing the beamforming vectors for interference alignment and closed-form solutions are obtained for some specific configurations. In contrast,  our work evaluates the necessary and sufficient conditions on channel structure of an IC to make perfect interference alignment possible with limited number of channel extensions.

\section{System Model} \label{sysmod}

\begin{figure}
\centering \includegraphics[scale=1.2]{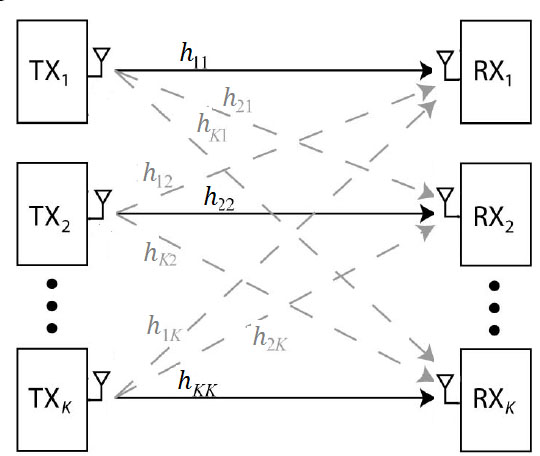}
\caption{K user IC Model.}
\label{figure:KUser}
\end{figure}

Consider the $K$ user IC consisting of $K$ transmitters and $K$ receivers each equipped with a single antenna,   as shown in Fig. \ref{figure:KUser}. Each transmitter wishes to communicate with its corresponding receiver. All transmitters share a common bandwidth and want to achieve the maximum possible  sum rate along with a reliable communication. Channel output at the $k^{\textrm{th}}$ receiver and over the $t^{\textrm{th}}$ time slot is characterized by the following
input-output relationship :
\begin{eqnarray}
{\bf Y}^{[k]}(t)={\bf H}^{[k1]}(t){\bf X}_p^{[1]}(t)+{\bf H}^{[k2]}(t){\bf X}_p^{[2]}(t) \cdots \nonumber \\+{\bf H}^{[kK]}(t){\bf X}_p^{[K]}(t)+{\bf Z}^{[k]}(t), 
\end{eqnarray}
where $k\in\{1,  \ldots,  K\}$ is the user index,   $t\in \mathbb{N}$ is the time slot index,   ${\bf Y}^{[k]}(t)$ is the output signal vector of the $k^{\textrm{th}}$ receiver,  ${\bf  X}_p^{[k]} (t)$  is the transmitted precoded signal vector of the $k^{\textrm{th}}$ transmitter which will shortly be defined,  ${\bf H}^{[kj]} (t)$,   $j\in\{1,  \ldots,  K\}$   is the fading factor of the channel from the $ j^{\textrm{th}}$ transmitter to the $k^{\textrm{th}}$ receiver  over  $t^{\textrm{th}}$ time slot,   and ${\bf Z}^{[k]}(t)$  is the additive white Gaussian noise at the $k^{\textrm{th}}$ receiver. The noise terms are all assumed to be drawn from a Gaussian independent and identically distributed random process with zero mean and unit variance. It is assumed that all transmitters are subject to  a power constraint $P$.

The channel gains are bounded between a positive minimum value and a finite
maximum value to avoid degenerate channel conditions (e.g. the case of all channel coefficients being equal or a channel coefficient being zero or infinite). Assume that the channel knowledge is causal and is available globally,   i.e. over the time slot $t$,   every node knows all channel coefficients ${\bf H}^{[kj]} (\tau),  \forall j,  k \in \{1,  \ldots,  K\},  \tau \in \{1,   \ldots,  t\}$. Hereafter,   time index is omitted for the sake of simplicity.

\section{Linear Interference Alignment Limitation}

Degrees-of-freedom region for a $K$ user IC, with the system model discussed in section \ref{sysmod}, has been derived in \cite{Cadam08} as follows, 
\begin{eqnarray}
\mathcal{D}= \left \{ {\bf d} \in \mathbb{R}_+^K: d_i+ d_j \leq 1, \; 1 \leq i, j\leq K \right \}, 
\label{dofreg}
\end{eqnarray}
and the number of DoF achieved by $K$ user IC is obtained to be $K/2$. Following corollay describes the only DoF vector, ${\bf d}$, that achieves total number of DoF. 

\begin{corollary}
\label{cor1}
The only DoF vector that achieves   total number of DoF of an IC  is
\begin{eqnarray}
d_i=\frac{1}{2}, \forall 1\leq i \leq K.
\label{optdof}
\end{eqnarray}
\end{corollary}
\begin{IEEEproof}
DoF point mentioned in (\ref{optdof}) is on the vertex of the DOF region. Since other vertices ${\bf e}_i$ has DOF less than $K/2$. Because DOF region is convex, it is straightforward to see that $d_i=\frac{1}{2}, \forall 1\leq i \leq K$ is the only point that produces the largest total DOF.
\end{IEEEproof}

Consider a three user IC. Assuming channel coefficients to be generic,   i.e. the channel coefficients are time varying and are drawn from continuous independent distributions,   \cite{Cadam08} has shown that optimal degrees of freedom for a three user IC cannot be achieved over limited number of channel usage. To maintain continuity of presentation, a short review is presented here. 

Consider using $2n$ time slots of the channel,   according to \ref{cor1}, achieving optimal degrees of freedom implies that $n$ degrees of freedom should be achieved for each of the transmitters. The signal vector at the $k$’th receiver can be stated as
\begin{eqnarray}
{\bf Y}^{[k]}&{}={}&{\bf H}^{[k1]}{\bf X}_p^{[1]}+{\bf H}^{[k2]}{\bf X}_p^{[2]}\nonumber \\ &&{+}\: {\bf H}^{[k3]}{\bf X}_p^{[3]}+{\bf Z}^{[k]},
\end{eqnarray}
where ${\bf X}_p^{[k]}$  is a $2n\times1$ column vector which is obtained by coding the transmitted symbols over $2n$ time slots of the channel,   as will be explained below. ${\bf Y}^{[k]}$ and ${\bf Z}^{[k]}$  represent the $2n$ symbol extension of $y^{[k]}$ and $z^{[k]}$,  respectively. ${\bf H}^{[kj]}$ is a diagonal $2n\times 2n$ matrix which represents the $2n$ symbol extension of the channel as shown in (\ref{topeq}) at the top of the next page.
\begin{figure*}[!t]
\normalsize
\setcounter{MYtempeqncnt}{\value{equation}}
\setcounter{equation}{4}
\begin{equation}
{\bf H}^{[kj]}\equiv \left [\begin{array}{c c c c} h^{[kj]}(2n(t-1)+1) & 0 & \cdots & 0\\ 0 & h^{[kj]}(2n(t-1)+2) & \cdots & 0 \\ \vdots & \cdots & \ddots & \vdots \\ 0 &  0  & \cdots & h^{[kj]}(2nt) \end {array}\right ]
\label{topeq}
\end{equation}
\setcounter{equation}{5}
\hrulefill
\vspace*{4pt}
\end{figure*}
In the extended channel,   message $ W_1$ at transmitter $1$ is encoded to $n$ independent streams $x_m^{[1]},   m=1,  \ldots,   n $  and sent along the vector ${\bf v}_m^{[1]} $ so that ${\bf X}_p^{[1]}$  can be written as 
\begin{eqnarray}
{\bf X}_p^{[1]} ={\bf V}^{[1]}  {\bf X}^{[1] },
\end{eqnarray}
where ${\bf X}^{[1]}$ is a $n\times 1$ column vector comprised of transmitted symbols  $x_m^{[1]},   m=1,  \ldots,   n $,   ${\bf V}^{[1]} $  is a $2n \times n$ dimensional precoding matrix comprised of the vectors  ${\bf v}_m^{[1]},    m=1,   \ldots,   n $ as its columns. In a similar way,   $W_2$ and $W_3$ are encoded by transmitters $2$ and $3$, respectively and sent to the channel as:
\begin{eqnarray}
{\bf X}_p^{[2]} ={\bf V}^{[2]}  {\bf X}^{[2] },
\end{eqnarray}
\begin{eqnarray}
{\bf X}_p^{[3]} ={\bf V}^{[3]}  {\bf X}^{[3] }.
\end{eqnarray}
The received signal at the $k'$th receiver can be evaluated to be 
\begin{eqnarray}
{\bf Y}^{[k]}&{}={}&{\bf H}^{[k1]}{\bf V}^{[1]}  {\bf X}^{[1] }+{\bf H}^{[k2]}{\bf V}^{[2]}  {\bf X}^{[2] } \nonumber \\ && {+}\:{\bf H}^{[k3]}{\bf V}^{[3]}  {\bf X}^{[3] }+{\bf Z}^{[k]}.
\end{eqnarray}

Receiver $i$ cancels the interference by zero forcing all ${\bf V}^{[j]},   j\not = i$ to decode $W_i$. At receiver $1$,   $n$ desired streams are decoded after zero forcing the interference from transmitters $ 2$ and $3$. To achieve $ n$ dimensions free of interference  from the $2n$ dimensional received signal vector ${\bf Y}^{[1]} $,  the dimension of the interference signal should not be more than $n$. This can be realized by perfectly aligning the received interference from transmitters $2$ and $3$ at the receiver $1$,   i.e.
\begin{eqnarray}
\textrm{span} \left ( {\bf H}^{[12]} {\bf V}^{[2]}  \right )=\textrm{span} \left ( {\bf H}^{[13]} {\bf V}^{[3]} \right ),
\label{SE1}            
\end{eqnarray}
where $\textrm{span}({\bf A})$ denotes the column space of matrix ${\bf A}$. At the same time, receiver $2$ zero forces the interference from ${\bf X}^{[1]}$  and ${\bf X}^{[3]}$. To achieve $n$ dimensions free of interference, we will have:
\begin{eqnarray}
\textrm{span} \left ( {\bf H}^{[21]} {\bf V}^{[1]}  \right )=\textrm{span} \left ( {\bf H}^{[23]} {\bf V}^{[3]} \right ).  
\label{SE2}           
\end{eqnarray}
In a similar way, ${\bf V}^{[1]}$  and ${\bf V}^{[2]}$  should be designed in a way to satisfy the following condition:
\begin{eqnarray}
\textrm{span} \left ( {\bf H}^{[31]} {\bf V}^{[1]}  \right )=\textrm{span} \left ( {\bf H}^{[32]} {\bf V}^{[2]} \right ).   
\label{SE3}           
\end{eqnarray}
Hence,  ${\bf V}^{[1]}$,   ${\bf V}^{[2]}$  and ${\bf V}^{[3]}$   should be chosen to satisfy (\ref{SE1}),   (\ref{SE2}) and (\ref{SE3}). Note that the channel matrices $ {\bf H}^{[ji]}$ are full rank almost surely. Using this fact,   (\ref{SE1}) and (\ref{SE2}) imply that 
\begin{eqnarray}
\textrm{span} \left (  {\bf V}^{[1]}  \right )=\textrm{span} \left ( {\bf T} {\bf V}^{[1]} \right ), 
\label{CAE1}           
\end{eqnarray}
where 
\begin{eqnarray}
 {\bf T}= ( {\bf H}^{[13]}  ) ^{-1} {\bf H}^{[23]}   ( {\bf H}^{[21]})^{-1} {\bf H}^{[12]} ( {\bf H}^{[32]}  ) ^{-1}  {\bf H}^{[31]}.
\label{TM}
\end{eqnarray}

If ${\bf V}^{[1]}$  could be designed to satisfy this criteria,   according to (\ref{SE2}) and (\ref{SE3}),   we can obtain ${\bf V}^{[2]}$  and ${\bf V}^{[3]}$  using
\begin{eqnarray}
 {\bf V}^{[2]} =\left ( {\bf H}^{[32]} \right )^{-1} {\bf H}^{[31]} {\bf V}^{[1]},
 \label{CAE2}    
\end{eqnarray}
\begin{eqnarray}     
  {\bf V}^{[3]} =\left ( {\bf H}^{[23]} \right )^{-1} {\bf H}^{[21]} {\bf V}^{[1]}.
\label{CAE3}           
\end{eqnarray}

 (\ref{CAE1}) implies that there is at least one eigenvector of  ${\bf T}$ in  $\textrm{span} \left (  {\bf V}^{[1]}  \right )$. Since all channel matrices are diagonal,   the set of eigenvectors for all channel matrices,   their inverse and product are all identical to the set of column vectors of the identity matrix,   i.e. the vectors of the from ${\bf e}_k=[0 \; 0 \; \cdots \; 1 \; \cdots \; 0]^T$. Since ${\bf e}_k$ exists in $\textrm{span} \left (  {\bf V}^{[1]}  \right )$,   (\ref{SE1})-(\ref{SE3}) imply that 
\begin{eqnarray}
&{}& {\bf e}_k \in \textrm{span} \left ( {\bf H}^{[ij]} {\bf V}^{[j]}  \right ),   \quad \forall i,   j \in \{1,  2,  3\}.         
\end{eqnarray}
Thus,   at receiver $1$,   the desired signal $ {\bf H}^{[11]} {\bf V}^{[1]} $  is not linearly independent of the interference signal,   ${\bf H}^{[12]} {\bf V}^{[2]}$,   and hence,   receiver $1$ cannot fully decode $W_1$ solely by zero forcing the interference signal. Therefore,   if the channel coefficients are completely random and generic,   we cannot obtain $3/2$ degrees of freedom for the three user single antenna IC through LIA schemes.

\section{Perfect Interference Alignment Feasibility Conditions}

In previous section,  If the objective was to align interference at two of the receivers,   receivers $1$ and $2$ for instance,   it could be easily attained using  (\ref{SE1}) and  (\ref{SE2}). Though,   as discussed above,    (\ref{SE3}) which refers to interference alignment criteria at receiver $3$,   cannot be satisfied simultaneously with  (\ref{SE1}) and  (\ref{SE2}). Instead,   assume channel matrices which contribute to interference at receiver $3$ would be of a form that already satisfies   (\ref{SE3}),   interference alignment would then be accomplished.

We can wait for the specific form of the channel to happen. The question we intend to answer in the following is that what is the necessary and sufficient condition on channel structure to make perfect interference alignment feasible in finite channel extension.

The following theorem summarizes the main result of this paper.

\begin{mytheo}
\label{3usertheo}
In a three user IC,  the necessary and sufficient condition for the perfect interference alignment to be feasible in finite channel extension is to have the following structure on the channel matrices:
\begin{eqnarray}
 &{}&{\bf T}= ( {\bf H}^{[13]}  ) ^{-1} {\bf H}^{[23]}   ( {\bf H}^{[21]})^{-1} {\bf H}^{[12]} ( {\bf H}^{[32]}  ) ^{-1}  {\bf H}^{[31]}\nonumber\\&&{=}\:{\bf P} \left [ \begin{array}{c c c} \tilde{{\bf T}} & 0 & 0 \\ 0 & \tilde{{\bf T}}& 0 \\ 0 & 0 & f(\tilde{{\bf T}}) \end{array} \right ] {\bf P}^T,
\label{CAEm}
\end{eqnarray}
where ${\bf P}$ is a $2n \times 2n$ permutation matrix,  $\tilde{{\bf T}}$ is an arbitrary $n_1 \times n_1$ diagonal matrix with nonzero diagonal elements,  and with $n_1$ in the range $1 \leq n_1 \leq n$,  and  $f({\bf X})$ is a  mapping whose domain is an arbirary $n_1 \times n_1$ diagonal  matrix and range is a $2(n-n_1) \times 2(n-n_1)$  diagonal matrix ${\bf Y}=f({\bf X})$ whose set of diagonal elements  is a subset of diagonal elements of  ${\bf X}$.
\end{mytheo}

\begin{remark}
Theorem \ref{3usertheo} simply states that matrix  ${\bf T}$ has no unique diagonal element.
\end{remark}

\begin{IEEEproof}
\begin{lemma}
\label{lemma1}
Assuming that $ {\bf V}^{[1]}$  is of rank $n$,   (\ref{CAE1}) implies that   $n$  eigenvectors of ${\bf T}$ lie in $\textrm{span} \left (  {\bf V}^{[1]}  \right )$.
\end{lemma}
\begin{IEEEproof}
From (\ref{CAE1}) we conclude that there exists a $n \times n$ dimensional matrix ${\bf Z}$ such that 
\begin{eqnarray}
{\bf T} {\bf V}^{[1]}={\bf V}^{[1]}{\bf Z}.
\end{eqnarray}
Assume that ${\bf u}$ is an eigenvector of ${\bf Z} $ i.e., ${\bf Z}{\bf u}=\gamma{\bf u}$ where $\gamma$ is its corresponding eigenvalue,   then ${\bf V}^{[1]} {\bf u} \not = 0$ and we can write:
\begin{eqnarray}
{\bf T} {\bf V}^{[1]}{\bf u}={\bf V}^{[1]}{\bf Z}{\bf u}=\gamma {\bf V}^{[1]}{\bf u}.
\end{eqnarray}
Then ${\bf V}^{[1]}{\bf u}$,   which is in $\textrm{span} \left ( {\bf V}^{[1]}  \right )$,   is an eigenvector of ${\bf T}$. Since  ${\bf Z}$ has $n$ orthogonal eigenvectors,   then $n$  orthogonal eigenvectors of ${\bf T}$ lie within $\textrm{span} \left ( {\bf V}^{[1]}  \right )$.
\end{IEEEproof}

$\textrm{span} \left ( {\bf V}^{[1]}  \right )$ should not contain any vector of the form ${\bf e}_i$,   and since $\textrm{span} \left ( {\bf V}^{[1]}  \right )$ has dimension $n$,   it should have $n$ basis vectors of the form $ {\bf v}\tilde{{\bf T}}=\sum_{i=1}^{2n}\alpha_i {\bf e}_i,   \quad j=1, \ldots,  n$,   where at least $2$ of $\alpha_i$'s are nonzero. Let's call vectors with this form as non ${\bf e}_i$ vectors. Since $n$ of  ${\bf T}$'s eigenvectors lie in $\textrm{span} \left ( {\bf V}^{[1]}  \right )$,   the matrix ${\bf T}$ should have at least $n$ non ${\bf e}_i $ eigenvectors. Note that this requirement is necessary not sufficient. Assuming that ${\bf S}=[{\bf s}]$ is a matrix consisted of non ${\bf e}_i $ eigenvectors of ${\bf T}$ as its columns, it is concluded that $\textrm{span} \left ( {\bf V}^{[1]}  \right ) \in \textrm{span} \left ( {\bf S}  \right )$.

\begin{lemma}
\label{lemma2}
${\bf T}$ has no unique diagonal element,  i.e.,  if $t_l$ is the $l$'th diagonal element of ${\bf T}$,  there is at least one $t_k, k=1, \ldots, 2n, k \neq l$ for which $t_l=t_k$.
\end{lemma}
\begin{IEEEproof}
It is easy to see that if ${\bf s}_1= {\bf e}_i + {\bf e}_j ,   \quad i,j=1,    \ldots,   n, i \neq j$  is an eigenvector of ${\bf T}$,   then $t_i = t_j$.   If $t_l$ is unique,  this implies that non ${\bf e}_i $ eigenvectors of ${\bf T}$ do not contain ${\bf e}_l$, and hence,  ${\bf e}_l \in \textrm{kernel} \left ( {\bf S}  \right )$, where $ \textrm{kernel} \left ( {\bf S}  \right )$  denotes  the null space of columns of matrix ${\bf S}$. Thus, $ {\bf e}_l \in \textrm{kernel} \left ({\bf V}^{[1]} \right )$ because  $\textrm{span} \left ( {\bf V}^{[1]}  \right ) \in \textrm{span} \left ( {\bf S}  \right )$. Since all channel matrices are diagonal,  using (\ref{CAE1})-(\ref{CAE3}),  ${\bf e}_j \in \textrm{kernel}({\bf V}^{[1]})$ implies that 
 \begin{eqnarray}
&{}& {\bf e}_j \in \textrm{kernel} \left ( {\bf H}^{[ij]} {\bf V}^{[j]}  \right ),    \quad \forall i, j \in \{1,   2,   3\}.
\end{eqnarray}

Thus,    at receiver $1$,    the total dimension of the desired signal $ {\bf H}^{[11]} {\bf V}^{[1]}$ plus interference from undesired transmitters, ${\bf H}^{[1j]} {\bf V}^{[j]},  j \neq 1$, is less than $2n$,  and desired signals are not linearly independent from the interference signals,    and hence,    receiver $1$ can not fully decode $W_1$ solely by zeroforcing the interference signal. 
\end{IEEEproof}
 
 Lemma \ref{lemma2} conlcludes the proof of necessary part of theorem \ref{3usertheo}. The sufficient part is easily proved by noting the fact that the matrix ${\bf T} $ with the form given in (\ref{CAEm}) has  $L \geq n$ non ${\bf e}_i $ eigenvectors ${\bf r}_i, i=1, \ldots, L$ with the property that 
 \begin{eqnarray}
{\bf e}_k \not \in \textrm{span}({\bf R}), \quad k=1, \ldots, 2n,
\label{spnprp}
\end{eqnarray}
 and
 \begin{eqnarray}
  {\bf e}_k \not \in \textrm{kernell}({\bf R}), \quad k=1, \ldots, 2n,
  \label{krnlprp}
  \end{eqnarray}
where ${\bf R} $  is defined as a $2n \times L$ matrix consisted of ${\bf r}_i$'s as its columns. Every $n$ subset of these eigenvectors can be considered as the columns of user $1$ transmit precoding matrix ${\bf V}^{[1]}$. ${\bf V}^{[2]}$   and ${\bf V}^{[3]}$ can be designed using (\ref{CAE2}) and (\ref{CAE3}).
\end{IEEEproof}

As an example, assume that, using $3$ extension of the channel,  $6 \times 6$  diagonal matrix  ${\bf T}$ has the following form, 
\begin{eqnarray}
{\bf T}=\textrm{Diag}(1, 2, 1, 2, 1, 2)
\label{Texpl}
\end{eqnarray}
which has the form given in (\ref{CAEm}) with $n=3, n_1=1$, ${\bf P}={\bf I}_6$, where  ${\bf I}_6$ is $6 \times 6$ identity matrix,  $\tilde{{\bf T}}=\textrm{Diag}(1, 2)$, and  $f(\tilde{{\bf T}})=\tilde{{\bf T}}$. The matrix ${\bf R}$ for this example case can be obtained as 
\begin{eqnarray}
{\bf R}=\left [  \begin{array}{c c c c} 1  & 1 & 0  & 0 \\ 0  & 0 & 1  & 1 \\ 
1  & 0 & 0  & 0 \\ 0  & 0 & 1  & 0 \\ 1  & -1 & 0  & 0 \\ 0  & 0 & 1  & -1 \end{array} \right ].
\end{eqnarray}
 Note that this choice for the set of non ${\bf e}_i$ eigenvectors of ${\bf T}$ defined in (\ref{Texpl}), which satisfies  (\ref{spnprp}) and (\ref{krnlprp}), is not unique. Every $6 \times 3$ matrix  ${\bf V}^{[1]} \in \textrm{span}({\bf R})$ can be considered as the user $1$ transmit precoding matrix. ${\bf V}^{[2]}$   and ${\bf V}^{[3]}$ can be obtained using (\ref{CAE2}) and (\ref{CAE3}). 
  
For the rest of the paper,   every matrix ${\bf U}$ which can be written in the form of (\ref{CAEm}),   with the same permutation matrix ${\bf P}$ and mapping function $f({\bf X})$,   would be stated as ${\bf U}={\bf T}_P$. It can easily be seen that if  ${\bf U}={\bf T}_P$ and  ${\bf V}={\bf T}_P$, so is  ${\bf U}^{-1}={\bf T}_P$  and  ${\bf U} {\bf V}={\bf T}_P$. 

\begin{remark}
If the condition (\ref{CAEm}) is true with the following form 
\begin{eqnarray}
 &{}&{\bf T}={\bf P} \left [ \begin{array}{c c} \tilde{{\bf T}} & 0 \\ 0 & \tilde{{\bf T}} \end{array} \right ] {\bf P}^T,
\label{CAEms}
\end{eqnarray}
where $\tilde{{\bf T}}$ is an an arbitrary $n \times n$ diagonal matrix, ${\bf V}^{[1]}$   can be designed as 
\begin{eqnarray}
{\bf V}^{[1]}={\bf P}^T \left[ \begin{array}{c} {\bf I}_n \\  {\bf I}_n \end{array} \right ], 
\label{vdesig}
\end{eqnarray}
where ${\bf P}$ is the same permutation matrix used in  (\ref{CAEms}) and ${\bf I}_n$ is the $n \times n$ identity matrix. ${\bf V}^{[1]}$ can also be designed as any other $2n \times n$ matrix having the same column vector subspace with (\ref{vdesig}). ${\bf V}^{[2]}$ and ${\bf V}^{[3]}$ are determined accordingly using (\ref{CAE2}) and (\ref{CAE3}), respectively. 
\end{remark}

\begin{remark}
Assuming channel aiding condition with the form given in (\ref{CAEms}),   consider the special case of ${\bf H}^{[ij]} ={\bf T}_P,   \quad \forall i,   j,   \quad i \not = j$,   then ${\bf T}={\bf T}_P$ and the channel aiding condition is already satisfied. The case of ${\bf H}^{[ij]} ={\bf T}_P,   i \not = j$  is the condition to satisfy the requirement of ergodic interference alignment in \cite{Nazer12},   therefore,   ergodic interference alignment is the special case  of the scheme presented in this paper.
\end{remark}

\section {Conclusion}
Channel aiding conditions obtained in this paper can be considered as the perfect Interference alignment feasibility conditions on channel structure. Stated conditions on channel structure are not exactly feasible,  assuming generic channel coefficients. Approximation can be used and its effect on residual interference can be analyzed. Overall,   this paper aims at reducing the required dimensionality and signal to noise ratio for exploiting degrees of freedom benefits of interference alignment schemes.

\bibliographystyle{IEEEtran}

\end{document}